\documentclass{elsart}

\usepackage{natbib,graphicx,amssymb}
\journal{New Astronomy}

\begin{document}

\begin{frontmatter}

\title{
MeV Gamma-Ray Imaging Detector with micro-TPC
}

\author[1]{T.~Tanimori},
%\ead{achieffi@ias.rm.cnr.it}
\author[1]{H.~Kubo},
\author[1]{K.~Miuchi},
\author[1]{T.~Nagayoshi},
\author[1]{R.~Orito},
\author[1]{A.~Takada},
\author[1]{A.~Takeda},
\author[1]{M.~Ueno},

\address[1]{Department of Physics, Kyoto University, Sakyo-ku, Kyoto
606-8502, Japan}

\begin{abstract}

We propose a new imaging gamma-ray
detector in the  MeV region.
By measuring the directions and energies of not only a scattered gamma ray 
but also a recoil electron, the direction of an incident gamma ray
would be essentially reconstructed event by event.
Furthermore, one of two measured (zenith and azimuth) 
angles of a recoil electron gives us 
an additional redundancy which enables us to
reject the  background events by kinematic constraints.
In order to measure the track of a recoil electron,
the micro Time Projection Chamber($\mu$-TPC) has been developed,
which can  measure the successive positions 
of the track of charged particles
in a few hundred micron meter pitch.  
The $\mu$-TPC consists of the new type of a gas proportional chamber:
micro PIxel gas Chamber ($\mu$-PIC) which is one of wireless gas
chambers and expected to be robust and stable.
Using this $\mu$-TPC and the Anger camera for the detection of  
a scattered gamma ray,  
we have obtained the first gamma-ray image by the full reconstruction
of the direction of gamma rays event by event.

\end{abstract}


\end{frontmatter}

\section{Introduction:}

Nowadays gamma-ray astronomy has become a very promising field of
astronomy.
However the MeV region of gamma rays is still uncultivated,
although this region would surely show us new aspects of high
energy phenomena in the universe.
In fact, COMPTEL in Compton Gamma-Ray Observatory had observed
for 9 years, and found about 30 celestial objects emitting
MeV gamma rays\citep{schon2002}.
However, taking into account of more than two hundreds discoveries 
of GeV gamma-ray emitters by EGRET, 
it is expected that 
more than a hundred celestial objects would
be detected in the MeV region. 
This is almost due to the nonexistence of a reliable imaging method  
of MeV gamma rays scattered via Compton process in the matter.   

\begin{figure}
\includegraphics*[width=0.5\linewidth]{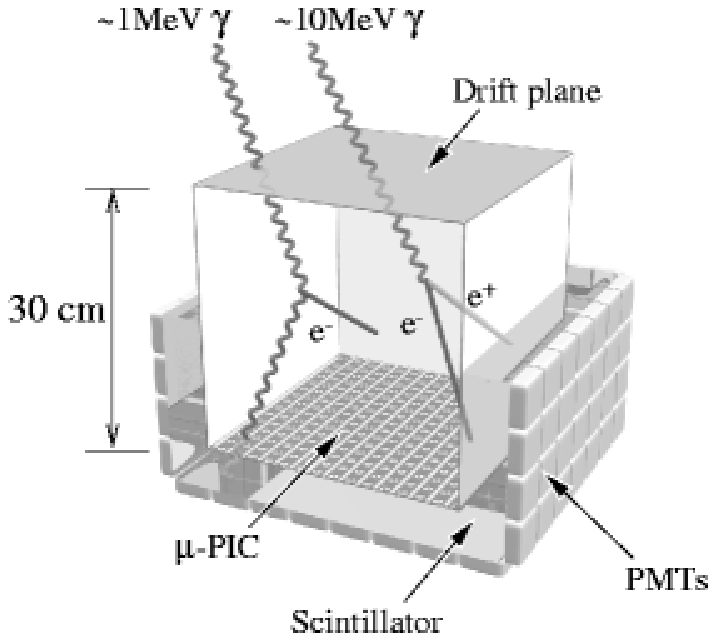}\hspace{6mm}
\includegraphics*[width=0.35\linewidth]{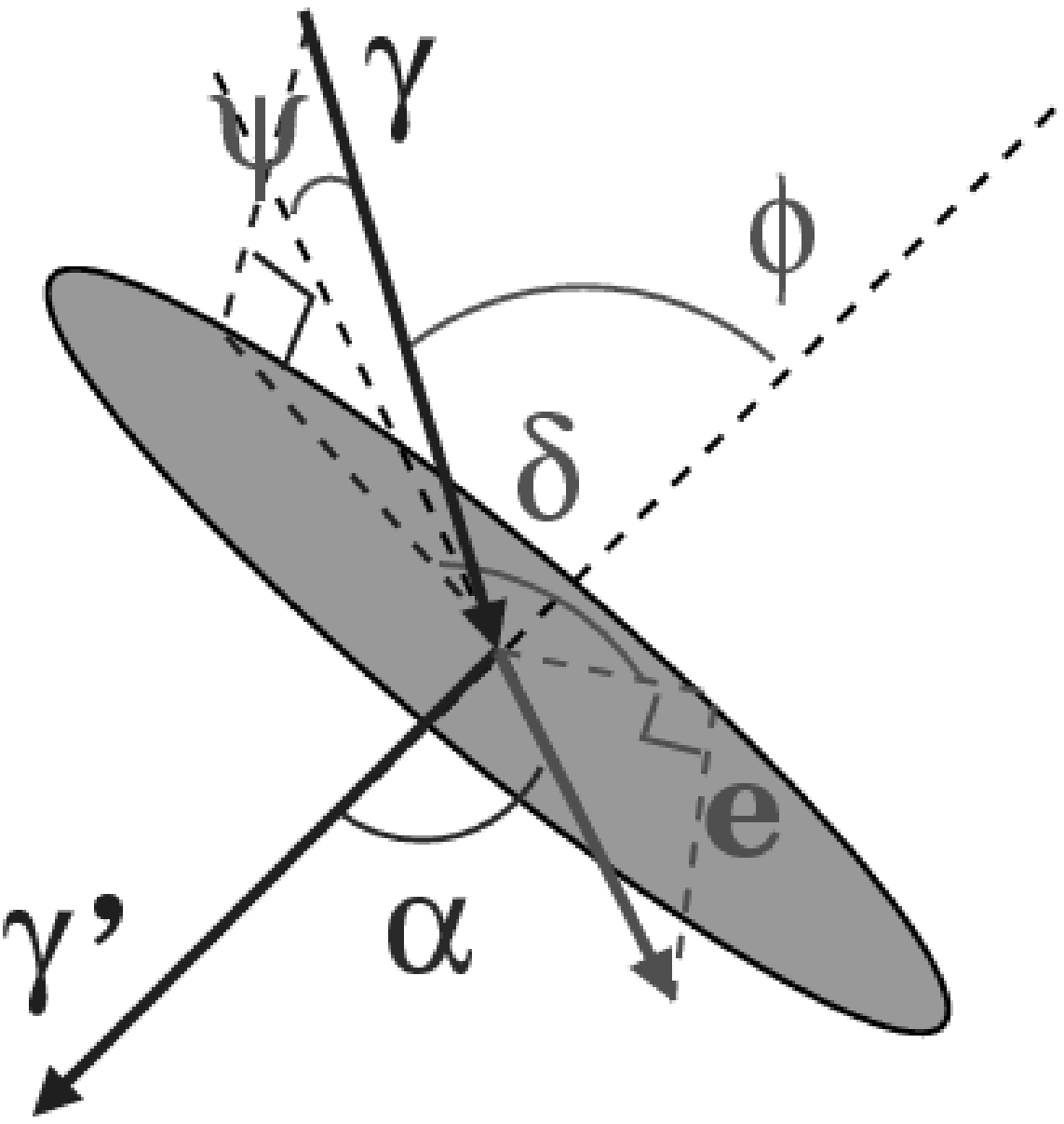}
\caption{
Schematic view of the planned detector for a balloon-borne experiment, 
which consists of 
a 30cm cubic $\mu$-TPC 
surrounded by a scintillation pixel detector(left figure).  
In the right figure, definitions of the 
angles are shown.}
\label{comp}
\end{figure}

Recently multi-Compton telescopes have eagerly been studied 
as a next generation of the detector launched in satellites\citep{Kurfess2003}.
Here we propose a new imaging detector of MeV gamma rays
measuring both the directions and energies of not only a scattered gamma ray 
but also a recoil electron,
which enables us to reconstruct the direction of an incident gamma ray
event by event.
Figure \ref{comp} shows the schematic view of the planned detector
based on this concept for a balloon-borne experiment, which consists of 
a 30cm cubic $\mu$-TPC 
surrounded by a scintillation pixel detector.
Also it is pointed out that  one of two measured 
angles of a recoil electron gives us 
an  additional redundancy which enables us to
reject almost all the  background events by the kinematic constraint.

Classical Compton telescopes measure 
the energies of both the scattered gamma ray and recoil electron,
and  one direction of the scattered gamma ray.
Since this method losses the direction of the recoil electron,
only one angle ($\phi$) of the incident gamma ray 
can be determined.
Furthermore there remains no redundancy to reject the background.
In the MeV region, not only strong celestial diffuse gamma rays 
but also the huge background gamma rays from the albedo and the satellite
itself by irradiation of cosmic rays come to the detector in space.
In addition, internal radioactivities in the detector emit both gamma rays
and beta decay electrons faking a real gamma ray in Compton detectors.
Good reviews of the analysis for the background in COMPTEL
were presented in this workshop\citep{Sch2003, Ryan2003}. 
In order to achieve a sensitivity ten times better than COMPTEL,
another powerful method for the background rejection seems to be inevitable
for the next  MeV gamma-ray imaging detector.

Here we introduce the new type of a
gamma-ray imaging detector based on a wireless gas detector, 
and the results of its performance study are presented.

\section{Detector}

In order to obtain a 3-D track of a recoil electron,
$\mu$-TPC has been developed,
which measures the successive positions of the track of charged particles
in a few hundred micron meter pitch\citep{Naga2003, Miuchi2003}.  
This $\mu$-TPC consists of the new type of a gas proportional chamber:
$\mu$-PIC which is one of wireless gas
chambers and expected to be more robust 
and stable than micro strip gas chambers\citep{Ochi2001, Ochi2002}.
At the first stage, 
we have developed a prototype  $\mu$-TPC having a detection volume of 
10 $\times$ 10 $\times$ 8 $\rm cm^3$.
All results mentioned here are obtained using this prototype detector.

\begin{figure}
\includegraphics*[width=0.4\linewidth]{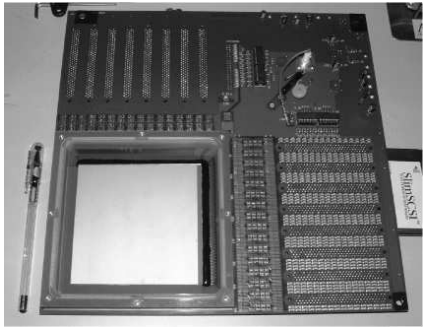}\hspace{10mm}
\includegraphics*[width=0.4\linewidth]{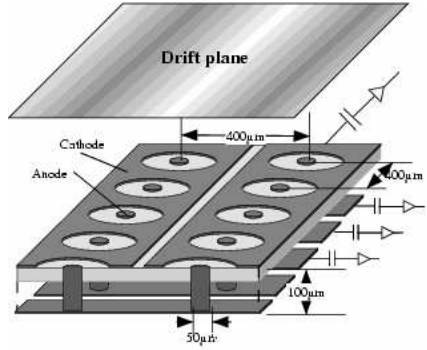}
\caption{
Top view of the 10cm square 2-D $\mu$-PIC attached on a PC-board (left),
and schematic view of the structure of $\mu$-PIC (right). }
\label{uPIC}
\end{figure}

\begin{figure}
\includegraphics*[width=0.4\linewidth]{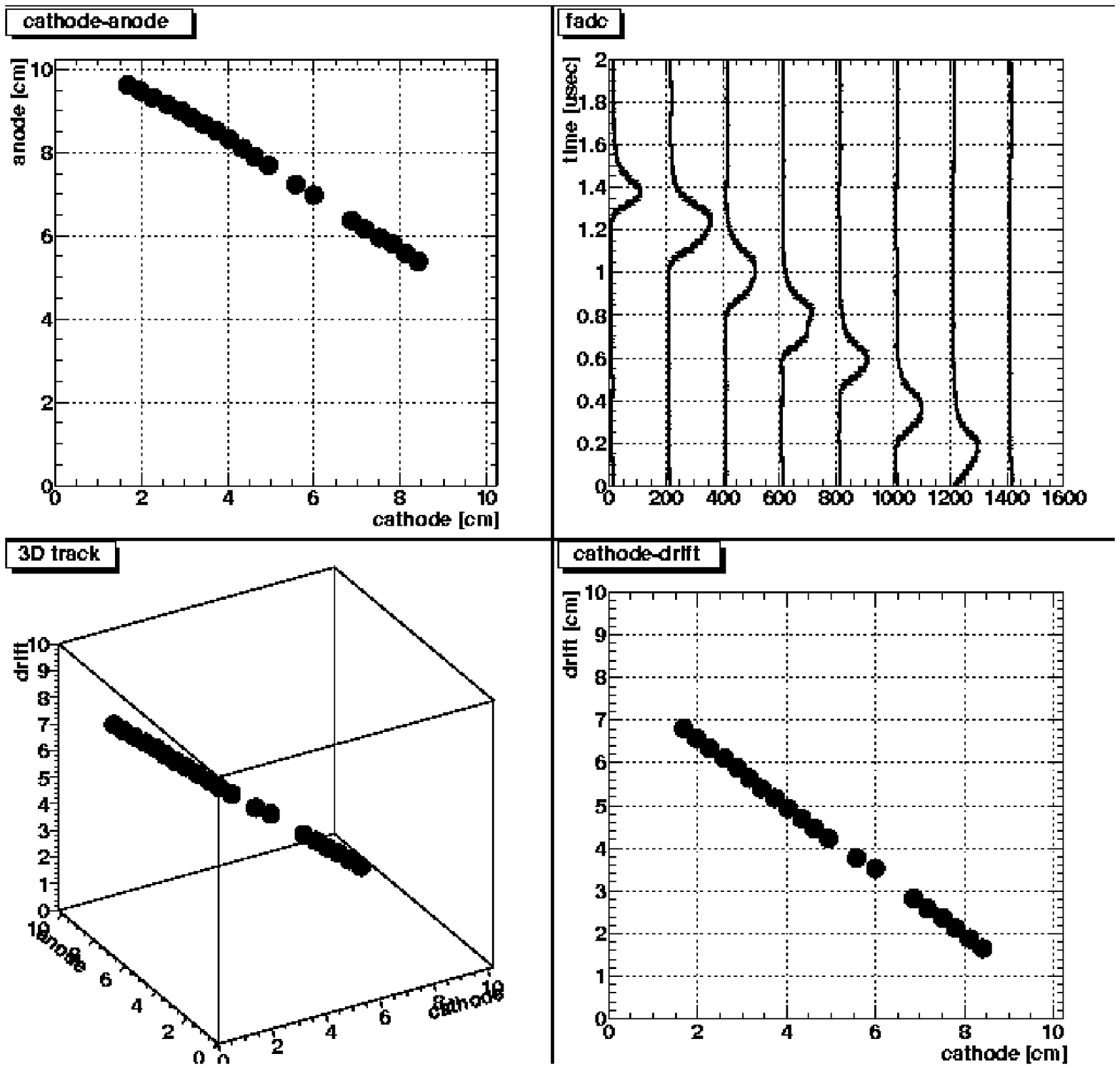}\hspace{10mm}
\includegraphics*[width=0.4\linewidth]{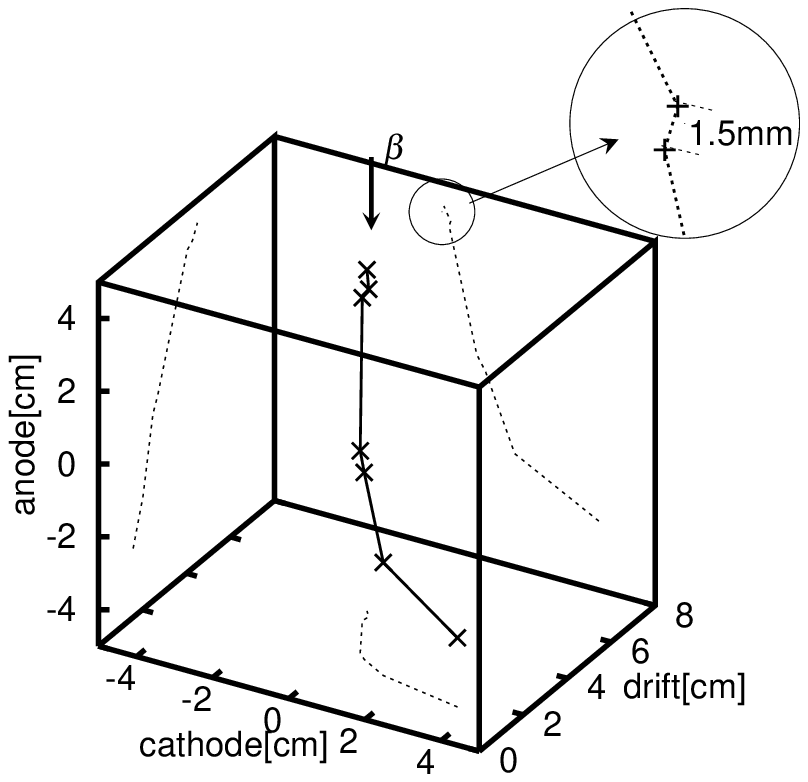}
\caption{Typical track of a 0.8 GeV proton (left)
%beam at KEK
and low energy electron emitted from $\beta$ decay (right) reconstructed
by $\mu$-TPC.}
\label{track}
\end{figure}

We briefly mention  the essential features of  
$\mu$-PIC and $\mu$-TPC.
$\mu$-PIC is a gaseous two-dimensional position-sensitive detector 
manufactured by the printed circuit board(PCB) technology. 
A schematic structure of the $\mu$-PIC is shown in Fig.\ref{uPIC}.
With the PC-board technology, 
large area detectors would be  mass-produced with low costs.
The pixel pitch of  the $\mu$-PIC is 400 $\mu$m. 
$\mu$-TPC was  stably operated with the  gas gain of $\sim$ 3000 for 
more than 1000 hours. 
Each signal of an anode and a cathode is amplified 
and shaped with 80ns time constant and discriminated.
For the energy measurement, 
analog signals of the amplifier connected to the cathode strips 
are summed 32 to one on the amplifier board, and 
its waveforms are recorded by the 8-channel 100 MHz flash ADC (FADC).
Energy resolution of $\sim 30 \%$(FWHM) was obtained for 5.9 keV X-rays.
Discriminated pulse (LVDS-level pulse) are fed to
the encoder board comprising five FPGA chips.
Here hit signals on the $\mu$-PIC are synchronized 
with an internal clock (20 MHz),
and their positions were recorded successively 
at an anode-cathode coincidence within one clock. 
Therefore tilted particle track in the drift volume 
can be recorded as successive hit points as shown in Fig.\ref{track}.
In this way, we realize both three-dimensional tracking and spectroscopy 
of the charged particles with $\mu$-TPC\citep{Kubo2003}. 
Three-dimensional spatial resolution of 260 $\mu$m was measured by 
the irradiation with proton beams.
This spatial resolution is restricted by the clock rate of the electronics
and the drift velocity in $\mu$-TPC ($\sim$ 5cm per 1$\mu$s). 
All the test on the $\mu$-TPC described here  were carried out 
using the Ar-$\rm C_2 H_6$ gas mixture with a normal pressure.

Fig. \ref{track} shows the examples of obtained tracks of low energy
proton (large {\it dE/dx}) and $\beta$-decay electron (near Minimum
Ionizing Particle: MIP), which indicates that
more gain ($\sim$10000, three times of the present gain)
is needed for the sufficient detection of MIPs.
The detection efficiency for MIPS is $\sim$ 10\% when requiring
at least 5 hit points in one track of MIP.
Now the electrode structure of $\mu$PIC is being improved.

\section{Imaging Performance}

\begin{figure}
\begin{center}
\includegraphics*[width=0.35\linewidth]{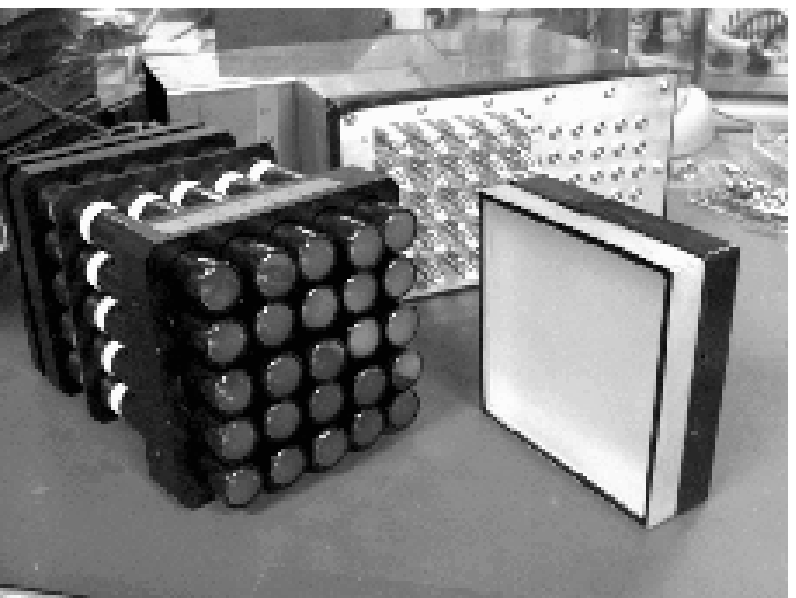}
\includegraphics*[width=0.5\linewidth]{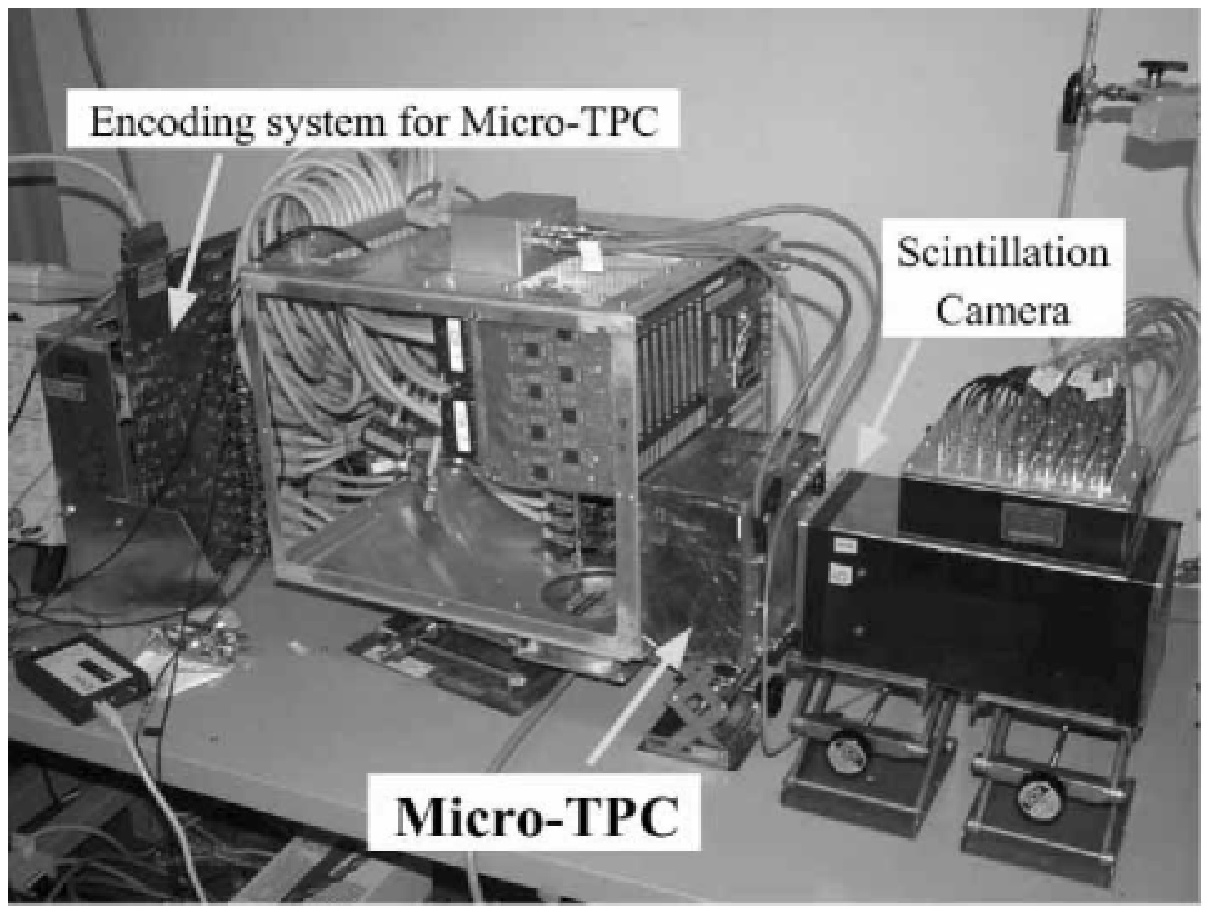}
\end{center}
\caption{
Photograph of the NaI(Tl) scintillator and 
the array of 3/4-inch PMTs(left), 
and side view of the set-up for the measurement of gamma-ray images,
where a gamma-ray source and Anger camera are placed 5cm behind the $\mu$-TPC
and 10cm at the front of the $\mu$-TPC, respectively(right).
}
\label{set-up}
\end{figure}

Scattered gamma rays are detected by 
the Anger scintillation camera consisting of
10 $\times$ 10 $\times$ 2.5 $\rm cm^3$ NaI(Tl) 
plate and 5 $\times$ 5 3/4-inch phototube array(Fig.\ref{set-up}).
Energy and position resolutions of the Anger camera 
for 662 keV gamma ray are 9 \% and 7.5mm at FWHM, respectively. 
Our gamma-ray imaging detector consists of $\mu$-TPC and the Anger camera,
as shown in Fig.\ref{set-up},
where a gamma-ray source and Anger camera are placed 5cm behind the $\mu$-TPC
and 10cm at the front of the $\mu$-TPC, respectively.
We examined the imaging performance of the prototype detector 
with three gamma-ray energies of $\rm {}^{133}Ba$(356keV), 
$\rm {}^{22}Na$(511keV), and $\rm {}^{137}Cs$(662 keV).
In the off-line analysis we required following constraints;
 the minimum number of hit positions
of the track in $\mu$-TPC (at least five points per track),
hit positions within  fiducial volumes of both $\mu$-TPC and the Anger camera,
a loose requirement of an energy deposit in  the Anger camera,
and the simple kinematic constraint  for $\alpha$ angle between
the scattered  gamma ray and recoil electron.
Since the energy calibration of the $\mu$-TPC was
not so accurate to sum with the energy of the Anger camera,
an initial energy of gamma rays was treated as a known parameter.

\begin{figure}
\begin{center}
\includegraphics*[width=0.35\linewidth]{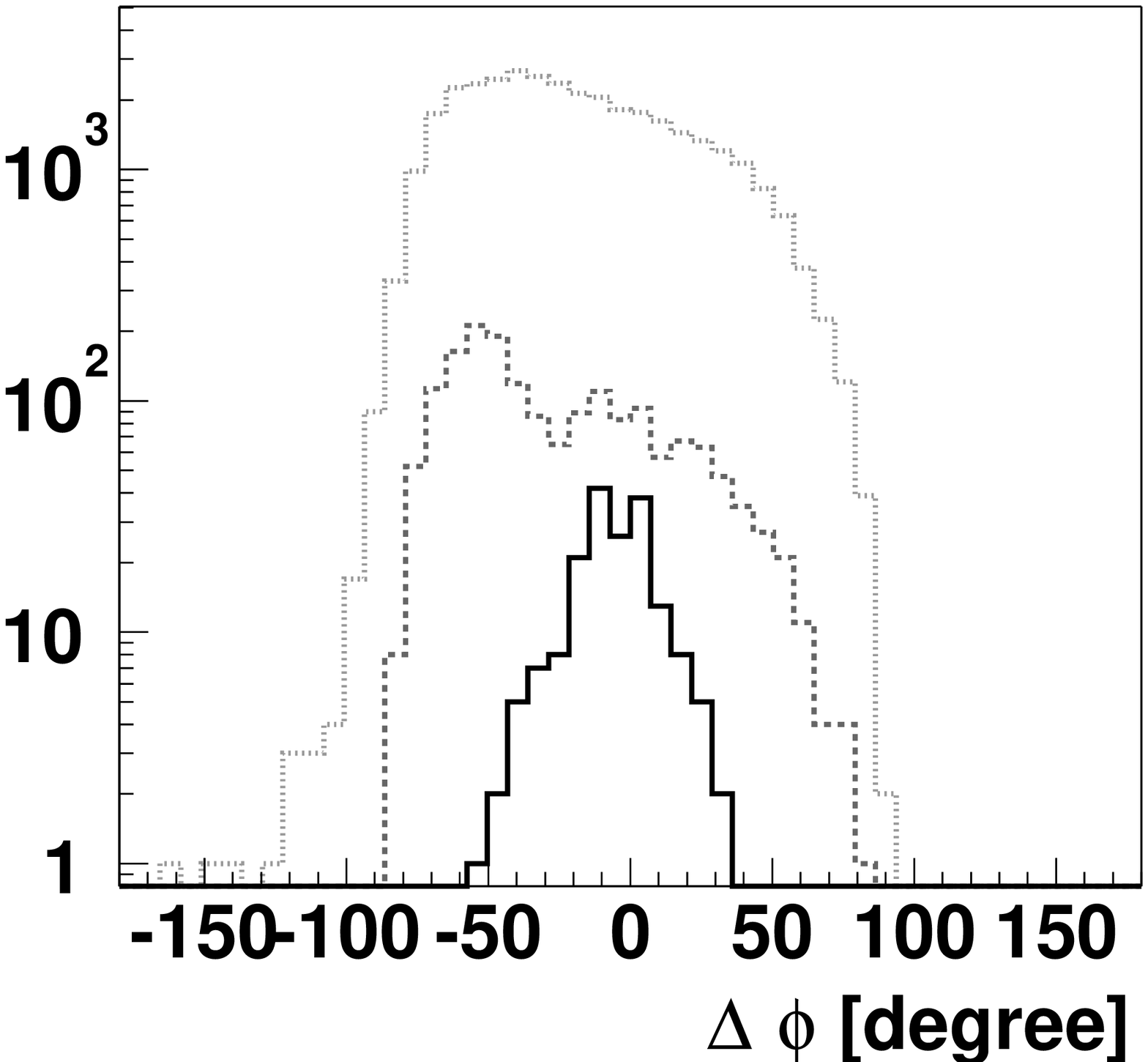}\hspace{7mm}
\includegraphics*[width=0.35\linewidth]{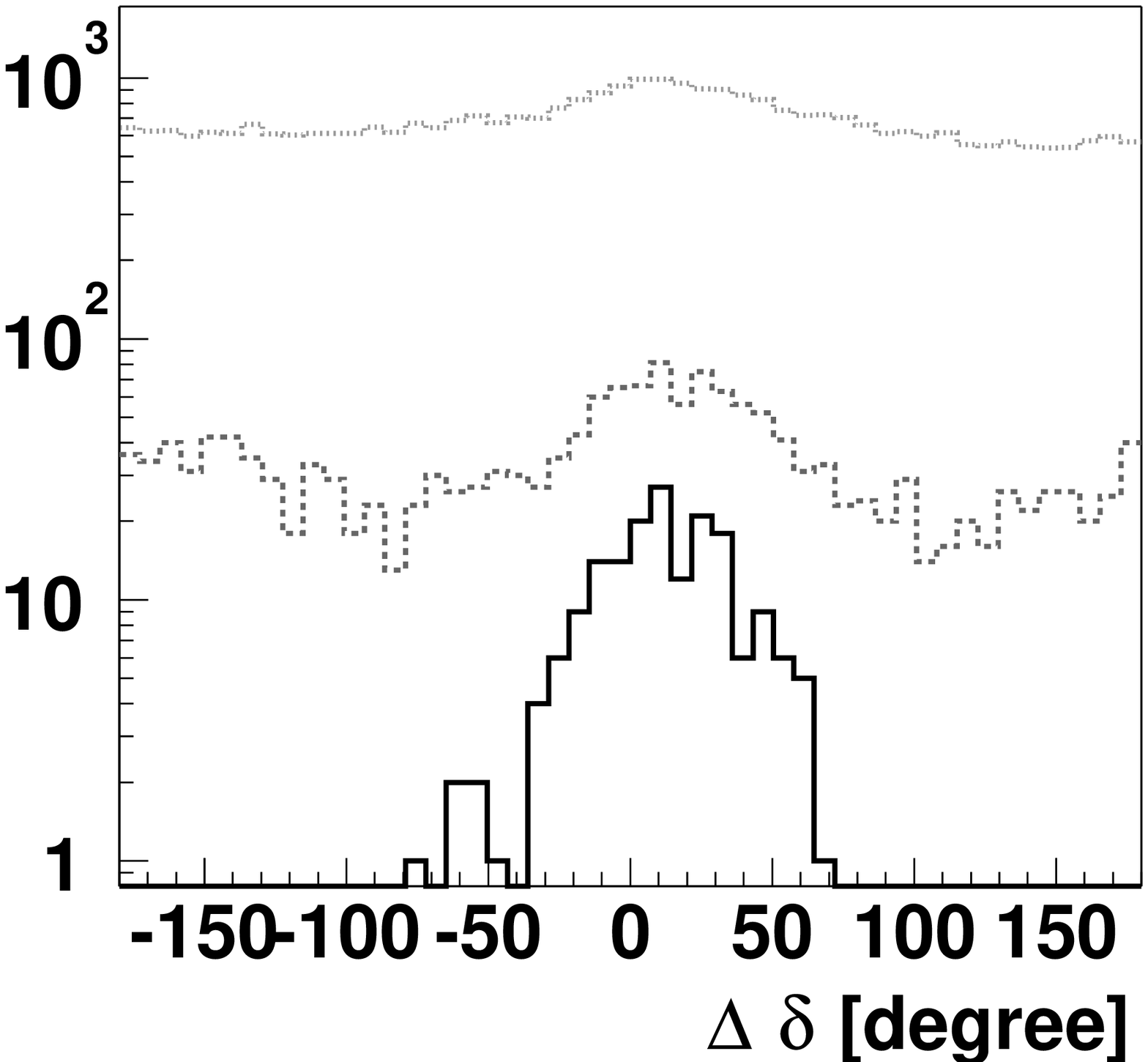}
\includegraphics*[width=0.4\linewidth]{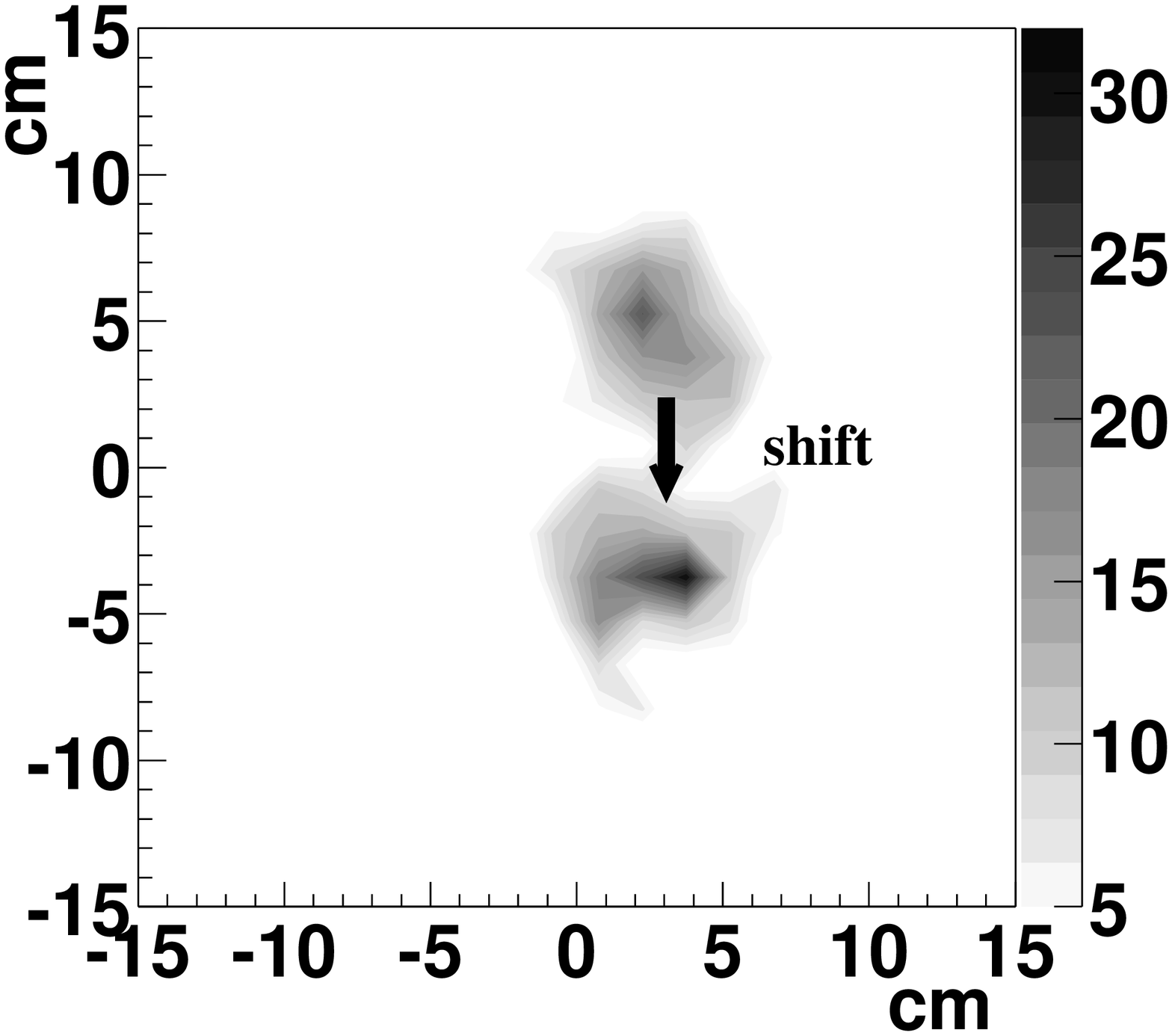}
\caption{
Angular resolutions of two angles of incident gamma rays,
$\phi$(top-left figure) and $\delta$(top-right figure),
where dotted-line, dashed-line and solid-one 
are obtained for raw data, 
after applying fiducial and energy cuts, and also after applying all-cuts,
respectively.
Also reconstructed images are plotted in the bottom figure,
where the source position was shifted about 40 degrees in the measurement, 
and  the reconstructed image was obviously also moved by this shift as shown
in the bottom figure.
}
\label{angular-dis.}
\end{center}
\end{figure}

Figure~\ref{angular-dis.} shows the  distributions of 
two angles of $\phi$ and $\delta$ obtained by the
event reconstruction
for $\rm {}^{137}Cs$(662 keV). 
Here we used only the events in which both a scattered gamma ray and a recoil
electron were detected, and hence we could determine the direction of
an incident gamma ray event by event.
We point out  that we did not use any imaging method such as a maximum entropy
method. 
Resultant  angular resolutions are 20 and 25 degrees at RMS,
respectively.
Also in Fig.~\ref{angular-dis.}, 
the reconstructed images of the gamma-ray source are presented,
where the source position was shifted about 40 degrees in the measurement, 
and  the reconstructed image was obviously also moved by this shift as shown
in the bottom figure.
The angular resolution of $\phi$, which is
used in classical Compton telescopes, is limited
by the position and energy resolutions of both detectors,
and our results  looks  near the expected one
from the energy and position resolutions of the Anger camera.
That of $\delta$ is limited by the multiple scattering
of the recoil electron and tracking resolution of $\mu$-TPC.
Obtained resolution of 25 degree is about three times worse
than the simulated one due to the insufficient gain of $\mu$-PIC.

Thus we have achieved the full reconstruction of Compton
process event by event  
and obtained the true gamma-ray image 
although the resolution of an image was still primitive.
Also  we point out that  a gas detector having a fine
tracking resolution such as $\mu$-TPC makes it possible 
to detect lower energy gamma rays from $\sim$ 200 keV.

\section{Simulation study for the balloon-borne experiment}

\begin{figure}
\includegraphics*[height=0.4\linewidth, width=0.5\linewidth]{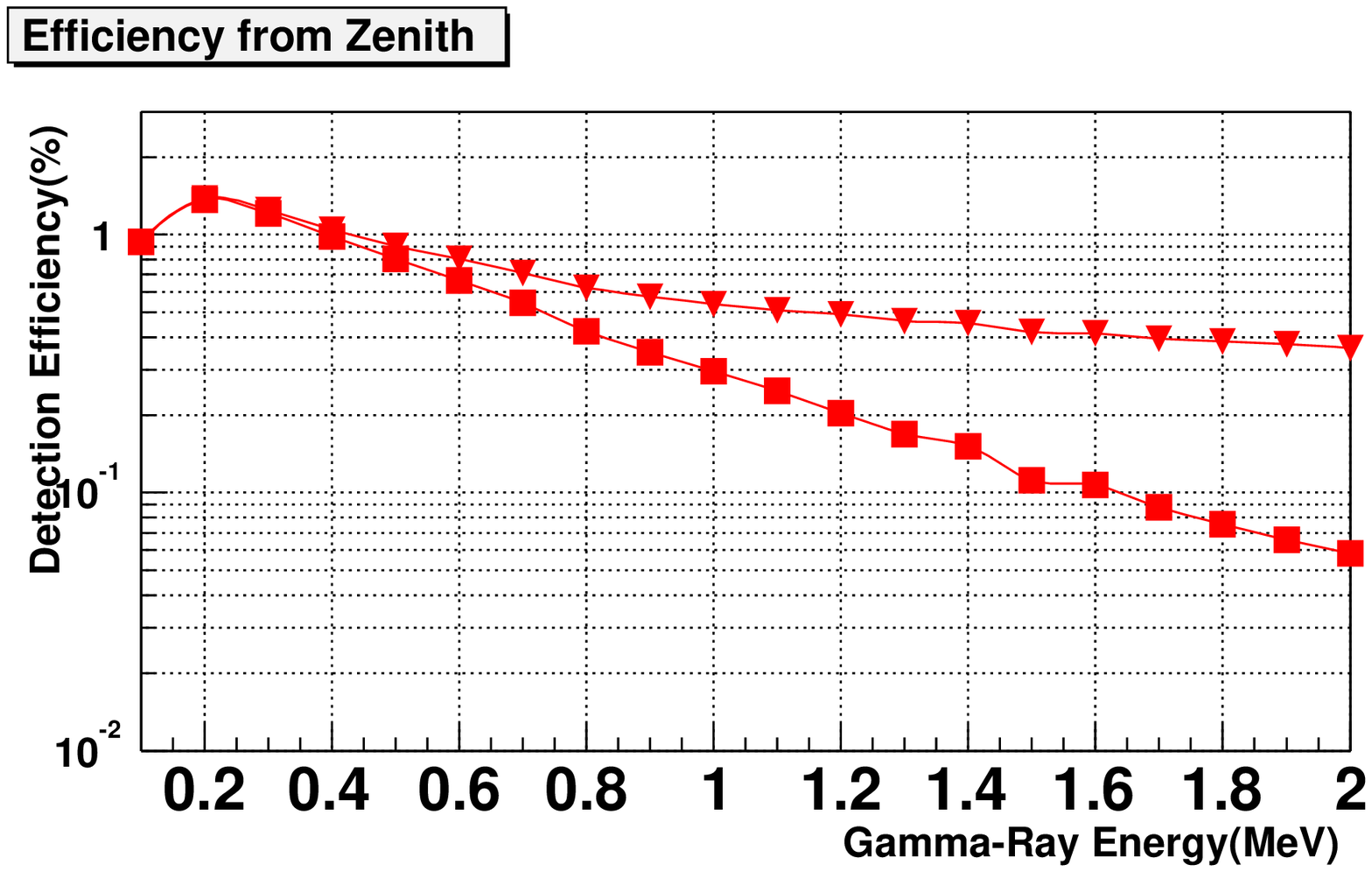}
\includegraphics*[height=0.4\linewidth, width=0.5\linewidth]{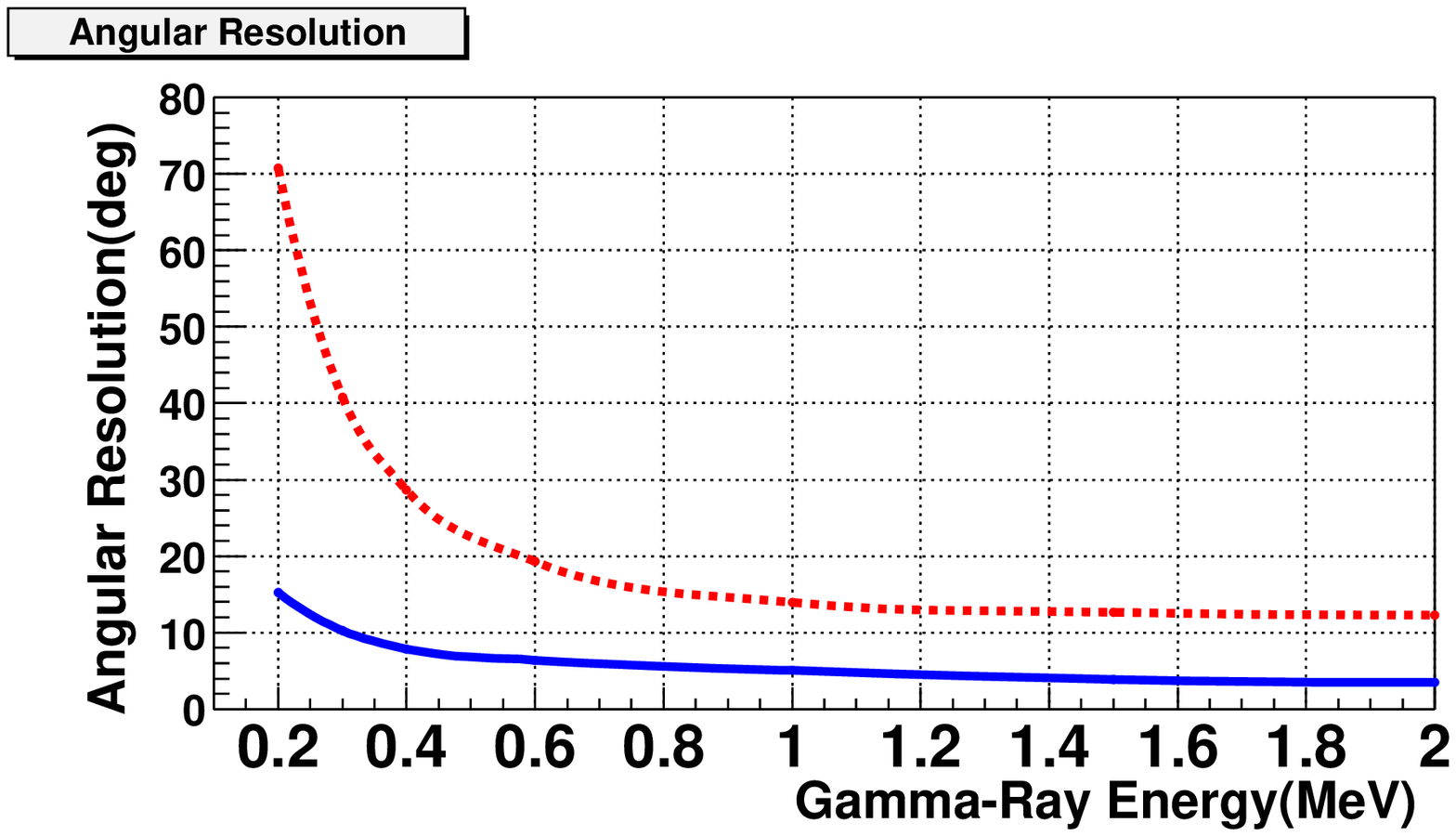}
\caption
{
Simulated detection efficiency (square plot in the left figure) and 
angular resolutions of two incident 
angles, $\delta$ and $\phi$(dotted and solid lines in the right figure)
as a function of the incident gamma-ray energy, 
where 1.5 bar pressured Xe gas in the $\mu$-TPC is assumed.
Although Ar-$\rm C_2 H_6$ mixture gas was used here for its convenience,
pressured Xe gas will be needed to obtain the high detection
efficiency for real observations.
In the left figure, triangle plots 
indicate the efficiency  when the energy of an electron passing
through the vessel were  obtained using a thin silicon pad detector
located out of the field cage of the TPC.
}
\label{simulation}
\end{figure}

We are developing this MeV gamma-ray imaging detector
for an all-sky survey in the MeV region,
and as a first step we plan to perform a balloon-borne
experiment launching a 30cm cubic $\mu$-TPC
surrounded by  a scintillation pixel detector 
(5mm square pixel assumed).
Simulated performances
for this detector are described in detail elsewhere\citep{Orito2003}.
For example, figure \ref{simulation}
shows the expected two angular resolutions
as a function of the gamma-ray energy.
This simulation study shows that a 30cm-cubic
detector has a  detection efficiency of $\sim$ 1 \% above
200 keV.
Assuming this efficiency with the use of 1.5 bar pressured Xe gas,
about one thousand gamma rays from the Crab 
in the energy region from 200 keV are expected to be detected
by 6 hours observation.

This work is supported by a Grant-in-Aid in 
Scientific Research of the Japan Ministry of 
Education, Culture, Science, Sports and Technology, 
and Ground Research Announcement for Space 
Utilization  promoted by Japan Space Forum.


\begin{thebibliography}{}

\bibitem[Sch$\ddot{\rm o}$nfelder et al. (2000)]{schon2002} 
Sch$\ddot{\rm o}$nfelder, V. et al. (2000), {\it A. \& AS.}, {\bf 143},~145

\bibitem[Kurfess (2003)]{Kurfess2003}
Kurfess, J., {\it this volume} 

\bibitem[Sch$\ddot{\rm o}$nfelder (2003)]{Sch2003}
Sch$\ddot{\rm o}$nfelder, V., {\it this volume} 

\bibitem[Ryan (2003)]{Ryan2003}
Ryan, J.~M., {\it this volume} 


\bibitem[Ochi et al. (2001)]{Ochi2001}
Ochi, A. et al. (2001), {\it Nucl. Instr. and Meth., A} {\bf 471}, 264

\bibitem[Ochi et al. (2002)]{Ochi2002}
Ochi, A. et al. (2002), {\it Nucl. Instr. and Meth., A} {\bf 478}, 196


\bibitem[Nagayoshi et al. (2002)]{Naga2003}
Nagayoshi, T. et al. (2003), to appear in {\it Nucl. Instr. and Meth., A}(hep-ex/0301008)


\bibitem[Kubo et al. (2003)]{Kubo2003}
Kubo, H. et al. (2003), to appear in {\it Nucl. Instr. and Meth., A} (hep-ex/0301009) 

\bibitem[Orito et al. (2003)]{Orito2003}
Orito, R. et al. (2003), to appear in {\it Nucl. Instr. and Meth., A}  

\bibitem[Miuchi et al. (2003)]{Miuchi2003}
Miuchi, K. et al. (2003), to appear in {\it IEEE Trans. Nucl. Sci.} 
{\bf 50},~(hep-ex/0301012)

\end{thebibliography}
\end{document}